# Controlling Metastability through Annealing of High-Entropy Nanoalloy Electrocatalysts to Boost Performance towards the Oxygen Evolution Reaction

Varatharaja Nallathambi[1,2*], Aneeta Jose Puthussery[3,4], Andrea M. Mingers[2], Robert Stuckert[1], André Olean-Oliveira[3,4], Sven Reichenberger[1], Dierk Raabe[2], Viktor Čolić[3,4], Baptiste Gault[2,5*], Stephan Barcikowski[1]

[1] Technical Chemistry I and Center for Nanointegration Duisburg-Essen (CENIDE), University of Duisburg-Essen, Universitaetsstr. 7, 45141 Essen, Germany

[2] Max Planck Institute for Sustainable Materials, Max-Planck-Str. 1, 40237 Düsseldorf, Germany

[3] Max Planck Institute for Chemical Energy Conversion, Stiftstr.34-36, 45470 Mülheim an der Ruhr, Germany

[4] CENIDE─Center for Nanointegration Duisburg-Essen, Carl-Benz-Str. 199, 47057 Duisburg, Germany

[5] Univ Rouen Normandie, CNRS, INSA Rouen Normandie, Groupe de Physique des Matériaux, UMR 6634, F-76000 Rouen, France

* corresponding authors: v.nallathambi@mpi-susmat.de, baptiste.gault1@univ-rouen.fr

**Abstract**

Low-cost transition metal high-entropy nanoalloys are emerging as sustainable alternatives to platinum group electrocatalysts. Synthesis conditions of single-phase solid solutions can alter phase stability, causing surface composition changes that affect electrocatalytic performance. Here, we propose to exploit the metastability of carbon-doped Cantor alloy-based amorphous high-entropy alloy nanoparticles produced by nanosecond-pulsed laser synthesis in organic solvents. *In situ* electron microscopy reveals crystallization and partitioning of elements upon heating to 600 °C, forming heterostructured nanoparticles with reinforced carbon shells that exhibit a 5- to 7-fold enhancement of the electrocatalytic activity compared to the as-synthesized counterparts for the oxygen evolution reaction. We demonstrate the strategic utilization of phase metastability in high-entropy nanoalloys through post-synthesis annealing to enhance the electrochemical activity of laser-generated nanoparticles.

**ToC Graphic**

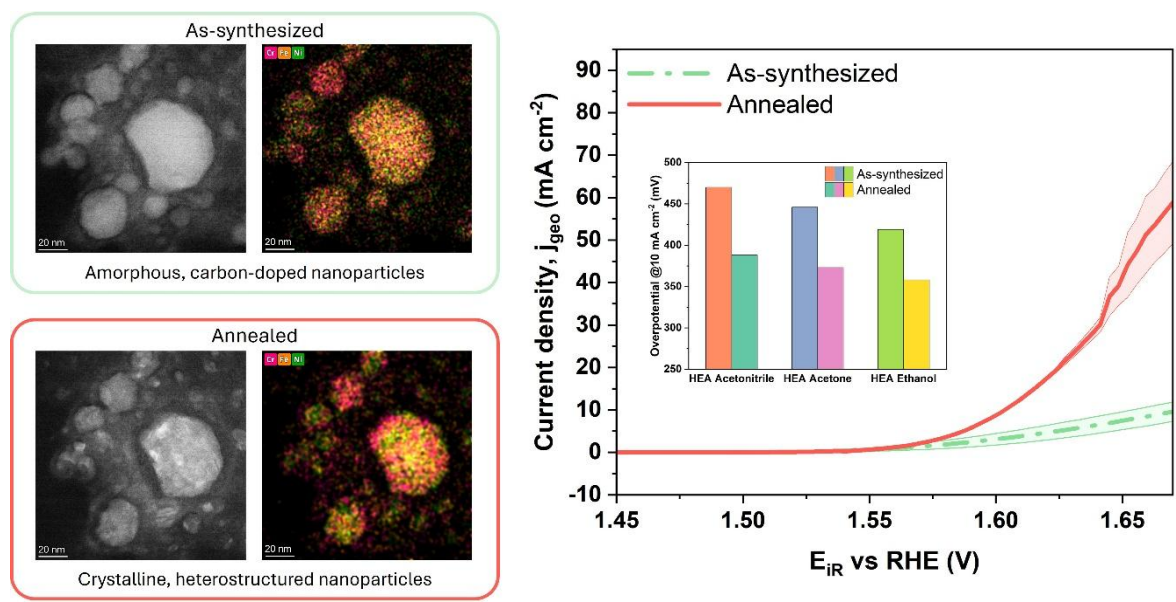

Achieving the sustainable development goals set for the coming decades requires innovative, experimental and strategic approaches to address critical needs in energy generation and storage [1,2]. A promising strategy involves substituting expensive platinum group electrocatalysts with abundant, sustainable transition metal-based systems to facilitate the upscaling of hydrogen generation [3,4]. Multicomponent systems known as high-entropy alloys (HEAs) or compositionally complex alloys (CCAs) that are transition metal-based were recently shown to exhibit electrocatalytic performance comparable to Pt [5–8], triggering considerable interest [9].

Large-scale, reliable synthesis of engineered HEA nanoparticles (NPs) was also demonstrated [10–15], particularly using laser synthesis and processing of colloids (LSPC) that involves laser ablation, fragmentation, and melting to fabricate functional NPs [16–22]. The extreme conditions during LSPC provide opportunities for design optimization [23–29], for crystalline and amorphous HEA NPs, attractive for catalytic applications [6,15,30–37]. We recently showcased control over the morphology, structure, and surface composition of HEA NPs through solvent selection during LSPC [38] of a Cantor alloy-based (CrMnFeCoNiC$_x$) HEA system synthesised in three solvents: acetonitrile, acetone, and ethanol. The laser-induced decomposition of solvent molecules resulted in doping of carbon (up to 20 at.%) into the HEA NPs. The supersaturation of carbon derived from solvent decomposition led to competitive dynamics between carbon shell formation and metallic coalescence and differences in NP characteristics [38]. The Cantor alloy is inherently metastable [39–44] and hence prone to phase separation and elemental partitioning [45,46]. In addition, the amorphous structure of the laser-generated HEA NPs resulting from carbon supersaturation is susceptible to crystallization at higher temperatures. [38].

Here, we strategically leverage the metastability to promote heat-induced, controlled phase separation in the HEA system, thereby generating heterostructured NPs with promising microstructures for electrocatalytic applications. Using *in situ* heating scanning transmission electron microscopy (STEM) coupled with energy dispersive X-ray spectroscopy (EDS) and X-ray photoelectron spectroscopy (XPS), we elucidate the temperature-dependent morphological changes, amorphous-to-crystalline transition, along with elemental partitioning within the core of the NPs and sub-surface layers beneath the carbon shell. The term 'high-entropy alloys' may not be entirely appropriate for describing phase-separated NPs, however, we use it to remain consistent with the literature. The oxygen evolution reaction (OER) activity performed at 1.65 V vs. the reversible hydrogen electrode (RHE) shows an increased catalytic activity up to 7-fold upon annealing, compared to the as-synthesized samples, which can be attributed to the interface-rich crystalline heterostructured NPs [9,47,48]. Scanning flow cell measurements coupled with online inductively coupled plasma–mass spectrometry (SFC–ICP–MS) [49–51] elucidate distinct dissolution behavior of the annealed samples that accounts for their enhanced catalytic activity. Our study demonstrates how manipulating phase metastability in high entropy nanoalloys can enhance catalytic performance. The annealing procedure presented here can be extrapolated to other synthesis methods involving complex multicomponent material systems, providing a new degree of freedom for engineering multifunctional high-entropy nanomaterials tailored for specific applications.

*In situ* STEM heating of HEA NPs synthesized in acetone allows for following the microstructural evolution during annealing (see Methods). The amorphous as-synthesized NPs, **Figure 1a**, have a rugged morphology, prominent carbon shells, and a near-uniform distribution of constituent elements in the NPs. At 400 °C, crystallization has initiated [32,38], yellow arrows in **Figure 1b**, and subtle changes in STEM-EDS elemental mapping indicate the onset of elemental partitioning. Upon heating to 600 °C, the amorphous-to-crystalline transformation is completed, accompanied by particle volume shrinkage and contrast variations arising from differently oriented lattice planes in **Figure 1c**. This is supported by powder X-ray diffraction (**Figure S1**). Elemental partitioning proceeds as individual Cr- and Ni-rich crystallites nucleate and grow, forming an interface-rich heterostructured microstructure, **Figure 1c** and **Figure S2**, that is interesting for electrocatalytic applications [9,47,48]. Ni-rich domains form preferably within the NP core, while Cr-rich domains appear both in the core and the subsurface layers. Co predominantly partitions with Ni, whereas Fe showed preferential accumulation in the core with sparse surface distribution. Mn shows a slight enrichment in Cr-rich regions, but predominantly migrates toward the surface and into the carbon shell, indicating the permeable nature of the carbon shell. Crystallization leads to C rejection from the NPs' core toward the shell (**Figure 1c**) [38]. Comparable observations were obtained across multiple sets of HEA NPs, **Figures S3 and S4**, as well as for NPs synthesized in acetonitrile (**Figures 1d, S5–S8**).

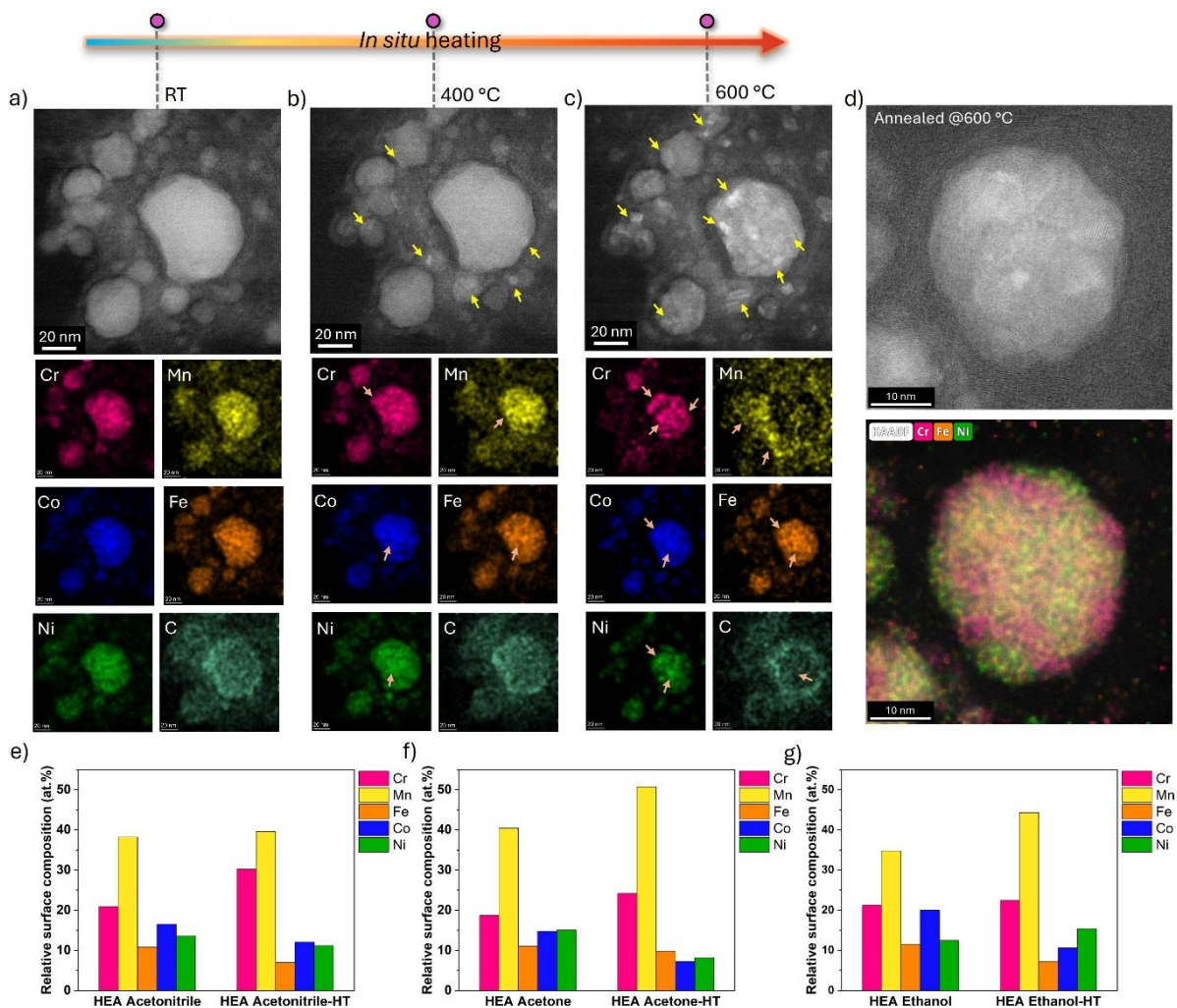

**Figure 1** *In situ* STEM heating results of HEA NPs synthesized in acetone. a), b), c) STEM dark field image and corresponding EDS mappings of the constituent elements at room temperature (RT), annealed at 400 °C and 600 °C, respectively; d) Examplary STEM dark field image and the corresponsing EDS overlay image of Cr, Fe and Ni combined highlighting the heterostructured microstructure following annealing at 600 °C of a HEA NP in acetonitrile; e), f), g) XPS relative surface composition analyses pre- and post-annealing heat treatment of the HEA NPs synthesized in acetonitrile, acetone and ethanol, respectively.

Annealing-induced surface compositional changes from XPS are plotted in **Figures 1e–g**. The as-synthesized NPs exhibit Mn surface enrichment (~ 38 at.%, 40 at.%, and 35 at.% for acetonitrile, acetone, and ethanol, respectively), while Cr distribution remained close to stoichiometric values. Fe, Co, and Ni were present below their expected stoichiometric ratios, consistent with our previous findings [38]. Following annealing at 600 °C, Mn and Cr are more highly concentrated on NP surfaces, which can be attributed to their preferential partitioning (**Figure 1c**). The detection of adventitious carbon [52,53] during XPS measurements complicates carbon quantification. While sputter cleaning or plasma treatment may reduce this interference, such approaches carry the inherent risk of inducing structural changes to the NP surface.

**Figure 2a** plots the cyclic voltammetry curves for HEA-NPs synthesized in three different solvents pre- and post-annealing. **Figure 2b** shows the linear sweep voltammetry curves for the OER at 1.65 V vs. RHE, the currents normalized with geometrical area ($j_{geo}$) obtained for the annealed samples synthesized in acetonitrile, acetone and ethanol were 21.15 mA cm$^{-2}$, 47.32 mA cm$^{-2}$ and 70.83 mA cm$^{-2}$, respectively, **Figure 2c**, which are approximately 5–7 times better performing when compared with the as-synthesized samples. Correspondingly, the overpotential values at 10 mA cm$^{-2}$ were substantially reduced for annealed samples: 388 mV, 373 mV, and 358 mV compared to 470 mV, 446 mV, and 419 mV for as-synthesized samples, **Figure 2d**. To further investigate the underlying electrochemical behavior differences, the electrochemically active surface area (ECSA) was determined via double-layer capacitance measurements at different scan rates. ECSA values increased by factors of 2.0, 2.3, and 1.5 for annealed samples as revealed by **Figure 2e**. The OER linear sweep voltammetry curves were also normalized by the ECSA determined from the double-layer capacitance, aligning with the trend obtained from the geometric-area normalization, suggesting that the observed activity correlates with intrinsic catalytic activity (**Figure S9**).

**Figure 2f** evidences a noticeable decrease that was observed for the solution resistance ($R_s$) values obtained from electrochemical impedance spectroscopy (EIS) using an equivalent electric circuit model (EEC, see Supporting Information) for all annealed samples, which can be attributed to the improved electrical contact and the adhesion between the substrate and the catalyst layer [54–59]. The charge transfer resistance ($R_{ct}$) revealed a significant decrease after annealing, implying enhanced electron transfer kinetics at the electrode-electrolyte interface. Specifically, the $R_{ct}$ values of the annealed samples in **Figure 2g** were reduced by 21%, 69%, and 82% compared to their as-synthesized counterparts (**Figure S10 and S11**). The Nyquist plots of the as-synthesized samples exhibited two distinct semicircles, which can arise from both the catalyst-electrolyte as well as the catalyst-substrate interface [56,60]. In contrast, the annealed samples exhibited a smaller semicircle, pointing out a simplified charge transfer process. This change can be from the improved crystallinity, thickened carbon shells, and annealing-induced reduction of interfacial defects, thereby increasing the conductivity [56]. The overall electrocatalytic performance improvement observed for the annealed samples arises from the microstructural and compositional transformations accompanying crystallization and heterostructure formation during annealing.

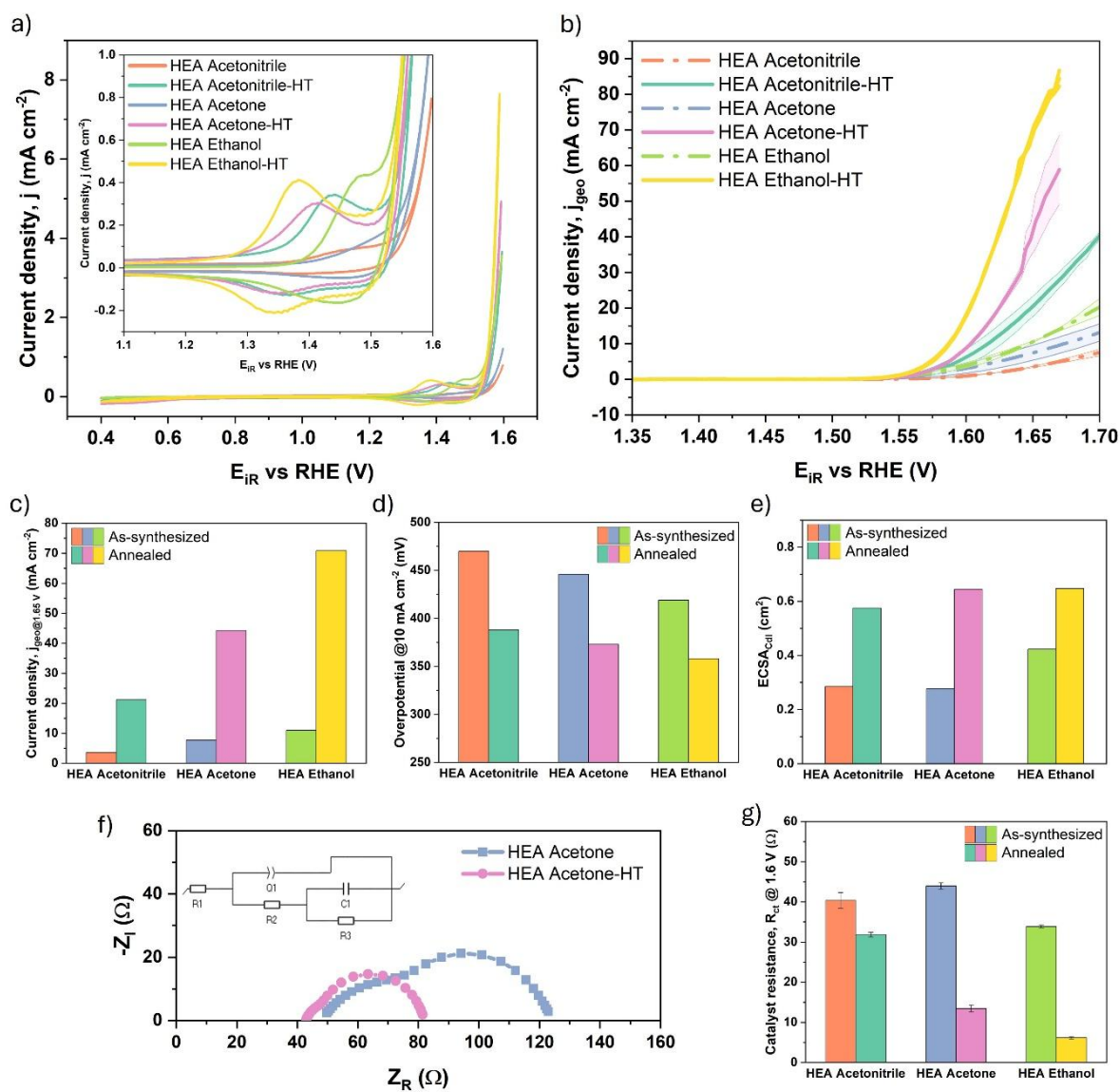

**Figure 2** Electrochemical characterization of the as-synthesized and annealed HEA NPs. a), b) Cyclic voltammetry and linear sweep voltametry curves (current density normalized with geometrical area) of the as-synthesized and annealed (denoted as HT) HEA NPs synthesized in acetonitrile, acetone and ethanol, respectively; c), d), e) Comparisons of current density measured @ 1.65 V vs. RHE, overpotential measured @ 10 mA cm$^{-2}$ and calculated ECSA of the as-synthesized and annealed HEA NPs synthesized in acetonitrile, acetone and ethanol, respectively; f) Exemplary Nyquist plots for the as-synthesized and annealed HEA NPs in acetone with the equivalent circuit for the spectrum given in the inset; g) Charge transfer resistance values measured at 1.6 V vs. RHE for as-synthesized and annealed HEA NPs in acetonitrile, acetone and ethanol, respectively.

The element-specific metal ion dissolution profiles obtained from SFC-ICP-MS measurements are presented in **Figures 3a** and **b** for as-synthesized and annealed HEA NPs in acetone. Differences in electrode contact area where the NP electrocatalysts are drop-cast with the SFC opening preclude direct comparison of dissolution concentrations across samples; however, trends can be meaningfully compared. Prominent contact dissolution peaks are observed when the electrode contacts the electrolyte under an applied potential of 1 V vs. the reversible hydrogen electrode (RHE) [38,46]. For as-synthesized samples, the highest dissolution concentration corresponds to Mn (42.9%), followed by Co and Ni (24% each), while Fe and Cr account for 5.1% and 4.0%, respectively. Despite the increased subsurface Cr concentration detected by XPS (**Figures 1e–g**), Cr dissolution is minimized by the protective carbon shells.

Post-annealing, the Mn dissolution concentration increases two-fold to 86.4%, while significant reductions are observed for all other elements: Co (6.9%), Fe (3.5%), Ni (2.5%), and Cr (0.8%). The increased Mn dissolution in annealed samples can be attributed to preferential Mn partitioning toward surface layers, as observed in STEM-EDS analysis (**Figures 1, S3–S8**), while heterostructure formation of Ni-Co-Fe-rich regions and reinforced carbon shells drastically reduces the dissolution of other elements compared to the as-synthesized state.

The dissolution behavior changes during potential sweeps from 1 V to 1.8 V vs. RHE, representing OER conditions. Cr dissolution begins with the onset of potential sweeping, followed by other elements, exhibiting an alternating pattern as a function of applied voltage, indicating periodic transitions between activated and deactivated states of the electrocatalyst surface. Near-constant dissolution of Fe, Co, and Ni occurs during potential sweeps for as-synthesized samples (**Figure 3a**), while annealed samples exhibit a regulated alternating pattern similar to Cr and Mn (**Figure 3b**). This represents a controlled dissolution of Fe, Co, and Ni under applied potential during OER, stemming from heterostructures formed and surrounded by Mn- and Cr-rich surface/subsurface layers beneath the carbon shell. The formation of Fe–Ni-rich oxyhydroxide surface layers following contact dissolution through a transient dissolution process during OER cycling has been reported previously [46,61]. In the case of as-synthesized material, after the initial dissolution of Cr and Mn from the surface, Fe, Co, and Ni from the subsurface region become exposed and subsequently form catalytically active oxyhydroxide phases. As the reaction progresses, these Fe–Co–Ni-rich surface layers may undergo exfoliation [46], after which Cr and Mn migrate toward the surface to replenish the dissolved species, allowing the cycle to repeat during continued operation.

In the annealed samples containing defined heterostructures and reinforced carbon shells, this transient dissolution and regeneration process occurs in a more controlled and spatially confined manner. The presence of pre-formed Fe-Co-Ni-rich heterostructures on the surface in a HEA matrix ensures the ready availability of multifunctional active sites, resulting in significantly enhanced catalytic activity [9]. The dissolution concentrations of Co and Ni in annealed samples are similar to those of Cr (**Figure 3b**), whereas they are considerably higher in as-synthesized counterparts (**Figure 3a**). This indicates the formation of more stable and accessible Ni-Co-Fe-rich active sites under applied potential in the heterostructured annealed samples. Analogous dissolution behaviors are observed in annealed NPs synthesized in acetonitrile and ethanol (**Figures S12 and S13**).

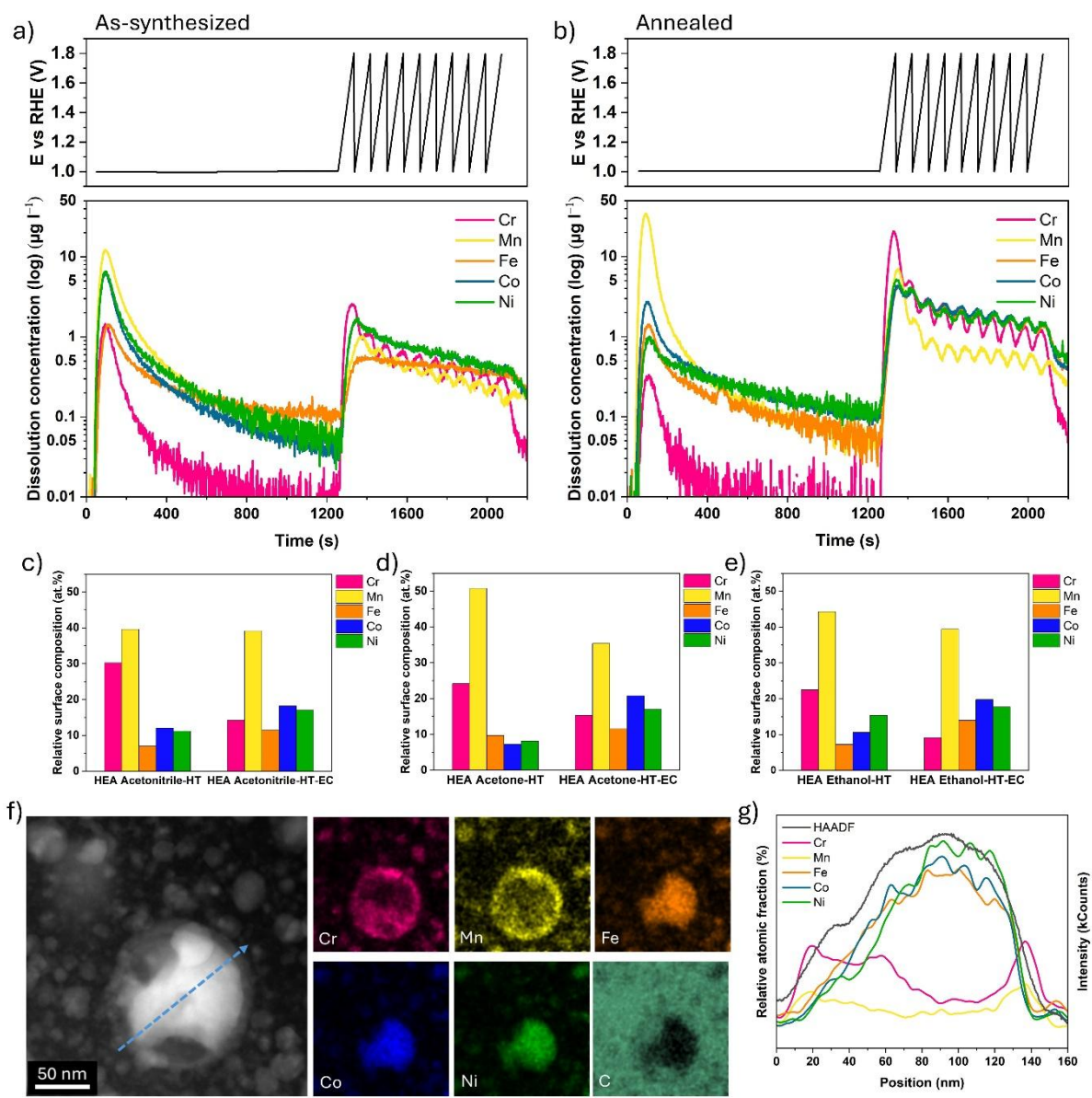

**Figure 3** Element-specific dissolution and compositional characterization of electrochemically tested annealed HEA NPs. a), b) Applied potential and the corresponding online ICP-MS dissolution profiles of dissolved metal ions as a function of time of the as-synthesized and annealed NPs in acetone, respectively; c), d), e) XPS relative surface composition analyses before and after OER cycling of annealed HEA NPs synthesized in acetonitrile, acetone and ethanol, respectively; f), g) STEM-EDS mapping of annealed HEA NPs synthesized in acetone after OER cycling and the corresponding one-dimensional composition profile, respectively.

Following OER cycling, XPS evidences a drop in the Mn and Cr surface concentration, **Figures 3c–e**, while the distribution of Fe, Co, and Ni increases, approaching near-equimolar ratios. Notably, surface Cr concentration was reduced by half following OER testing compared to the annealed state, which relates to the higher Cr dissolution observed during the beginning of potential sweeps in SFC-ICP-MS (**Figure 3b**). The relative surface concentrations of Fe, Co, and Ni became comparable across all three samples after OER cycling, indicating a steady-state surface elemental distribution under reaction conditions accompanied by the transient dissolution process. It should be noted that the microstructural and compositional transformations observed during *in situ* heating TEM experiments may not fully represent *ex situ* annealing in a furnace. To bridge this gap, STEM analysis was conducted on annealed HEA NPs following electrochemical testing. STEM-EDS mapping presented in **Figure 3f** reveals the elemental distribution in annealed samples after OER testing, showing a surface layer within the carbon shell enriched in Cr and Mn that encapsulates a heterostructured core containing Fe, Co, and Ni. Cr-rich regions are also observed near the core, adjacent to Ni-rich domains. **Figure 3g** presents the one-dimensional compositional profile across the NP. These observations substantiate the trends observed in the dissolution profiles of the annealed sample (**Figure 3b**). **Figure S14** shows additional STEM-EDS mapping of a large nanoparticle (~320 nm in diameter), further confirming the retention of the heterostructured microstructure after OER testing.

In summary, we leveraged the thermodynamic metastability of high-entropy alloy NPs from pulsed-laser synthesis in three different organic solvents to enhance their electrocatalytic performance through post-synthesis annealing. Annealing triggers multiple microstructural modifications: (i) metastable carbon-containing amorphous NPs crystallize through the formation of fine crystallites within individual particles, accompanied by a volume shrinkage; (ii) carbon supersaturation is reduced by ejection of carbon into the surrounding shell; (iii) elemental partitioning leads to NPs featuring Cr-Fe-rich and Ni-Co-Fe-rich heterostructured regions; (iv) Mn partitions to the NPs' subsurface and shell. These heterostructured annealed NPs exhibit significantly enhanced electrochemical performance. Given their comparable surface compositional distributions following electrochemical testing, the performance enhancement can be primarily attributed to the formation of numerous interfaces within Cr-Fe-rich and Ni-Co-Fe-rich crystalline heterostructured microstructures and the controlled elemental dissolution under reaction conditions in a carbon shell-protected, subsurface-enriched environment, achieving a maximum activity enhancement of 7-fold. We demonstrated microstructural engineering in multicomponent nanoalloys by exploiting their inherent metastable nature, showcasing their potential for optimized electrocatalytic performance.

**Materials and methods**

*Target preparation*

The equimolar CrMnFeCoNi high-entropy alloy (HEA) target was fabricated through vacuum induction melting and casting of pure metallic elements. The as-cast alloy target was then subjected to a hot and cold rolling procedure to attain the desired thickness. The target was then homogenized at 1200 °C for 3 hours under an argon atmosphere before water quenching. A final annealing treatment was carried out at 900 °C for 1 hour in an argon atmosphere, followed by water quenching.

*Pulsed-laser ablation in liquids*

Nanoparticle synthesis was achieved through pulsed-laser ablation utilizing an EdgeWave GmbH nanosecond laser system operating at 1064 nm wavelength with 8 ns pulse duration, 5 kHz repetition frequency, and 35 mJ pulse energy. The polished bulk target was placed within a flow chamber through which nitrogen-purged organic solvents (acetonitrile, acetone, or ethanol) circulated continuously at a flow rate of approximately 50 mL min$^{-1}$.

*Annealing heat treatment*

Post-synthesis heat treatment of the nanoparticles in powder form was performed using a heating furnace (Carbolite Gero Ltd., England). The thermal treatment protocol consisted of two stages: initial pre-conditioning at 400 °C for 1 hour (heating rate: 5 °C/min), followed by heating to 600 °C for 10 minutes (heating rate: 1 °C/min) in a 10% $H_2$ in Ar atmosphere. An identical thermal treatment procedure was applied to glassy carbon electrodes with dropcast colloidal nanoparticles prior to electrochemical testing.

*Materials characterization*

Powder X-ray diffraction (XRD) measurements of nanoparticles pre- and post-annealing was conducted using a Rigaku Smartlab 9 kW diffractometer equipped with Cu Kα radiation (λ = 0.15406 nm). The diffraction patterns were collected over a 2θ range of 20° to 120° using a step increment of 0.01° and a scanning rate of 1°/min. The measurements were performed under operating conditions of 45 kV accelerating voltage and 200 mA tube current.

Morphology and compositional characterization of the nanoparticles were carried out using a probe-corrected Thermo Fisher Titan Themis 300 scanning transmission electron microscope (STEM) equipped with energy dispersive X-ray spectroscopy (EDS) capabilities. The instrument was operated in STEM mode at 300 kV acceleration voltage, 100 mm camera length, and 23.8 mrad beam convergence angle. Imaging was conducted using bright field, annular dark field and high-angle annular dark field (HAADF) detectors. Image acquisition and EDS data collection were performed using Thermo Fisher Scientific's Velox 3.6.0 software. Elemental mapping and quantitative analysis focused on the Kα emission lines of Cr, Mn, Fe, Co, and Ni. Quantitative EDS analysis employed the standard Cliff-Lorimer (K-factor) method with absorption corrections. Each EDS map represented an accumulation of a minimum of 100 frames at 1024 × 1024 pixel resolution, with a minimum pixel dwell time of 10 μs to achieve adequate spectral intensity for reliable analysis. A kernel-based redistribution function (pre-filtering) within the Velox software was used to improve the quality of the EDS data before quantification.

X-ray photoelectron spectroscopy (XPS) analysis was conducted to determine the relative surface elemental composition of the nanoparticles using a Ulvac-Phi VersaProbe II system. The instrument used Al-Kα radiation at 1486.6 eV with a 100 μm spot size and 0.5 eV energy resolution. Data acquisition employed a hemispherical analyzer positioned at 45° to the sample surface, dual-beam charge neutralization, and a pass energy of 23 eV. Sample preparation involved drop-casting the nanoparticles onto glassy carbon substrates to enable characterization under various conditions: as-synthesized, annealed, and after electrochemical testing. Spectral analysis and peak deconvolution were performed using CasaXPS software with Shirley background [62] applied to individual peak fits. Energy calibration of high-resolution XPS spectra was referenced to the adventitious carbon peak at 284.8 eV, determined through C 1s spectrum deconvolution for each sample [52]. Peak fitting employed symmetric Gauss-Lorentzian line shapes, with detailed fitting constraints for the 3p spectra of all elements and corresponding peak fits provided in the Supporting Information. Surface composition quantification was based on peak areas derived from the deconvoluted 3p signals.

*In situ heating studies*

In-situ heating experiments were conducted using a probe-corrected Thermo Fisher Titan Themis 300 transmission electron microscope operating at 300 kV. The experiments employed a specialized in-situ chip holder featuring amorphous SiNx windows (Nano-Chips Wildfire Double Tilt – DENSsolutions), onto which the colloidal nanoparticles were deposited via drop-casting and subsequently air-dried. Plasma cleaning of the sample holder with the chip in place was carried out before analysis. The heating protocol involved a stepwise temperature increase to 600 °C, with 100 °C increments starting from 300 °C at a heating rate of 1 °C/s and a hold time of 10 minutes to ensure thermal equilibrium. Imaging and EDS acquisition were carried out after cooling down the stage to room temperature (RT) after each temperature step.

*Electrochemical testing*

Electrochemical characterization was performed using a three-electrode glass cell configuration with a VSP-3e potentiostat (BioLogic, France). The electrode setup consisted of a custom-fabricated reversible hydrogen electrode (RHE, reference), a Pt wire counter electrode (Mateck, 99.99% purity), and a rotating disk electrode featuring a glassy carbon disk with 0.196 cm$^2$ surface area (Pine Research; AFMSRCE, USA) as the working electrode. The electrolyte solution was prepared by dissolving potassium hydroxide pellets (Sigma-Aldrich Germany, 85% purity) in ultrapure water (Merck Millipore, Milli-Q IQ 7003, 18.2 MΩ) to achieve a 0.1 M KOH concentration. Sample preparation involved drop-casting HEA nanoparticle colloids onto the glassy carbon electrode surface, followed by drying under a nitrogen atmosphere. The volumes of colloidal solutions drop-cast onto the glassy carbon electrode surfaces were calculated based on colloidal concentration to ensure uniform loading of 10 μg cm$^{-2}$ across all samples. Prior to drop-casting, the glassy carbon electrode underwent sequential polishing with 1 μm, 0.3 μm, and 0.05 μm alumina particles on microcloth, followed by 5-minute sonication in ultrapure water. Charge transfer resistance measurements were obtained through electrochemical impedance spectroscopy (EIS) performed at 1.6 V vs RHE, scanning from 30 kHz to 1 Hz with 10 mV amplitude and 6 step dec$^{-1}$. Electrode cleanliness was verified through cyclic voltammetry of the bare glassy carbon electrode. Sample conditioning involved 50 cyclic voltammetry cycles at 50 mV s$^{-1}$ between 0.4 V and 1.6 V until stable responses were achieved. Following a 30-minute argon purging of the electrolyte and with electrode rotation at 1600 rpm, linear sweep voltammetry measurements representing oxygen evolution reaction (OER) were conducted from 1.3 V to 1.7 V at a scan rate of 10 mV s$^{-1}$. iR drop compensation was performed by a combination of real-time measurement as well as during the data analysis. 85 % of the uncompensated resistance was corrected via the Z-impedance-resistance (ZIR) technique and the remaining 15 % was compensated during data processing.

*Scanning flow cell measurements*

Elemental dissolution studies using electrochemical measurements were conducted in a micro-electrochemical scanning flow cell (SFC) [49,50] fabricated from polycarbonate (Makrolon) and using a Gamry Reference 600 potentiostat. The cell configuration featured a Pt-wire as counter electrode (0.5 mm diameter, 99.997% purity, Alfa Aesar) positioned in the inlet channel and an Ag/AgCl/3 M KCl reference electrode located in the outlet channel, with both channels having a diameter of 1.9 mm. The electrolyte solution of 0.01 M $H_2SO_4$ was prepared from suprapure $H_2SO_4$ (96%, Merck) and ultrapure water (PureLab Flex2, Elga, 18 MΩ cm$^{-1}$, TOC < 3 ppb). Electrolyte circulation was maintained at approximately 380 μL min$^{-1}$ through the cell before direct introduction into an inductively coupled plasma mass spectrometer (ICP-MS, NexION 300X, Perkin Elmer) for real-time quantification of dissolved metal ions.

A calculated volume of colloidal nanoparticle solutions to maintain a loading of 1 μg was dropcast onto a polished glassy carbon plate serving as the working electrode. To prevent any detachment of the catalyst during measurements, 1 μL of Nafion/isopropanol solution (20 mL/1000 mL was dropcast on top of them. The working electrode was then positioned to be enclosed by the SFC opening.

A four-point calibration of the ICP-MS was performed every day prior to the measurements to convert detected intensities to dissolved ion concentrations ($^{52}$Cr, $^{55}$Mn, $^{56}$Fe, $^{59}$Co, $^{60}$Ni). An internal standard ($^{74}$Ge at 50 μg L$^{-1}$ in 0.01 M $H_2SO_4$) was used to correct for changes in analyte intensities due to physical interferences, which was introduced via a Y-connector behind the SFC. Time scale synchronization was carried out for the approximately 45-second delay (ranging from 43 to 52 seconds) between ion dissolution at the working electrode and ICP-MS detection. The electrochemical protocol initiated with potentiostatic conditioning at 1 V vs. reversible hydrogen electrode (RHE) to establish electrode contact under potential control. Following subsidence of the initial contact dissolution peak, ten consecutive potential ramps from 1 V to 1.8 V vs. RHE were executed at a 10 mV s$^{-1}$ scan rate.

**Supporting Information**

Supporting Information is available from ACS Publications or from the author.

**Acknowledgments**

S.B. acknowledges funding by the Deutsche Forschungsgemeinschaft (DFG) - Project 277627168. V.N. is grateful for the financial support from the International Max Planck Research School for Interface Controlled Materials for Energy Conversion (IMPRS-SurMat), now International Max Planck Research School for Sustainable Metallurgy (IMPRS-SusMet), and the Center for Nanointegration Duisburg-Essen (CENIDE). The authors are grateful to Philipp Watermeyer and Volker Kree for their support at the TEM facilities at the Max Planck Institute for Sustainable Materials. Further, the authors acknowledge Benjamin Breitbach (Max

Planck Institute for Sustainable Materials) for performing the XRD measurements. V. N. is grateful to Peter Schweizer and Siyuan Zhang at the Max Planck Institute for Sustainable Materials for the helpful discussions on *in situ* STEM measurements. The authors thank the Interdisciplinary Center for Analytics on the Nanoscale (ICAN) at the University of Duisburg-Essen (UDE) for access to the XPS facility. V.N. and R.S. thank Ulrich Hagemann at the ICAN for the help with XPS measurements. V.N. greatly appreciates Aparna Saksena and Yongqiang Kang at the Max Planck Institute for Sustainable Materials for the helpful discussions regarding the scanning flow cell experiments.

**Declaration of competing interest**

The authors declare that they have no known competing financial interests or personal relationships that could have appeared to influence the work reported in this paper.

**Data availability statement**

The data that support the findings of this study are available from the corresponding authors upon reasonable request.

# Controlling Metastability through Annealing of High-Entropy Nanoalloy Electrocatalysts to Boost Performance towards the Oxygen Evolution Reaction

Varatharaja Nallathambi[1,2*], Aneeta Jose Puthussery[3,4], Andrea M. Mingers[2], Robert Stuckert[1], André Olean-Oliveira[3,4], Sven Reichenberger[1], Dierk Raabe[2], Viktor Čolić[3,4], Baptiste Gault[2,5*], Stephan Barcikowski[1]

[1] Technical Chemistry I and Center for Nanointegration Duisburg-Essen (CENIDE), University of Duisburg-Essen, Universitaetsstr. 7, 45141 Essen, Germany

[2] Max Planck Institute for Sustainable Materials, Max-Planck-Str.1, 40237 Düsseldorf, Germany

[3] Max Planck Institute for Chemical Energy Conversion, Stiftstr.34-36, 45470 Mülheim an der Ruhr, Germany

[4] CENIDE─Center for Nanointegration Duisburg-Essen, Carl-Benz-Str. 199, 47057 Duisburg, Germany

[5] Univ Rouen Normandie, CNRS, INSA Rouen Normandie, Groupe de Physique des Matériaux, UMR 6634, F-76000 Rouen, France

* corresponding authors: v.nallathambi@mpi-susmat.de, baptiste.gault1@univ-rouen.fr

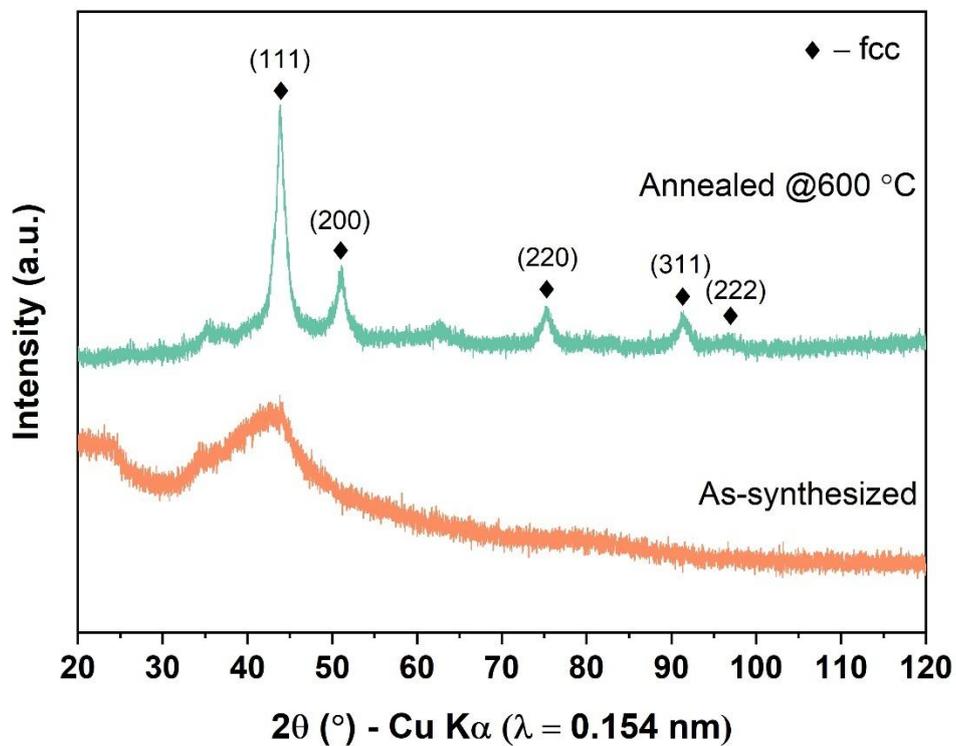

Figure S1 XRD patterns of as-synthesized amorphous and annealed crystalline HEA NPs. Strong reflections of an fcc phase can be observed for the annealed sample with minor reflections at ~35° and ~63° which could be attributed to surface oxides.

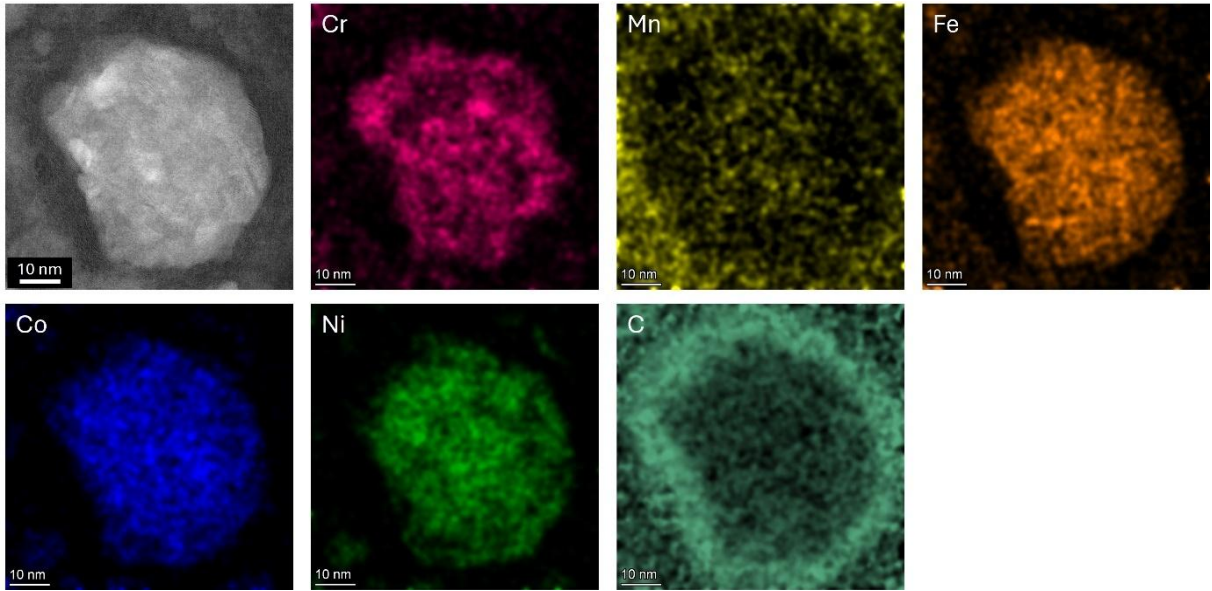

Figure S2 STEM dark field image and corresponding EDS mappings of the constituent elements of a HEA NP synthesized in acetone annealed at 600 °C.

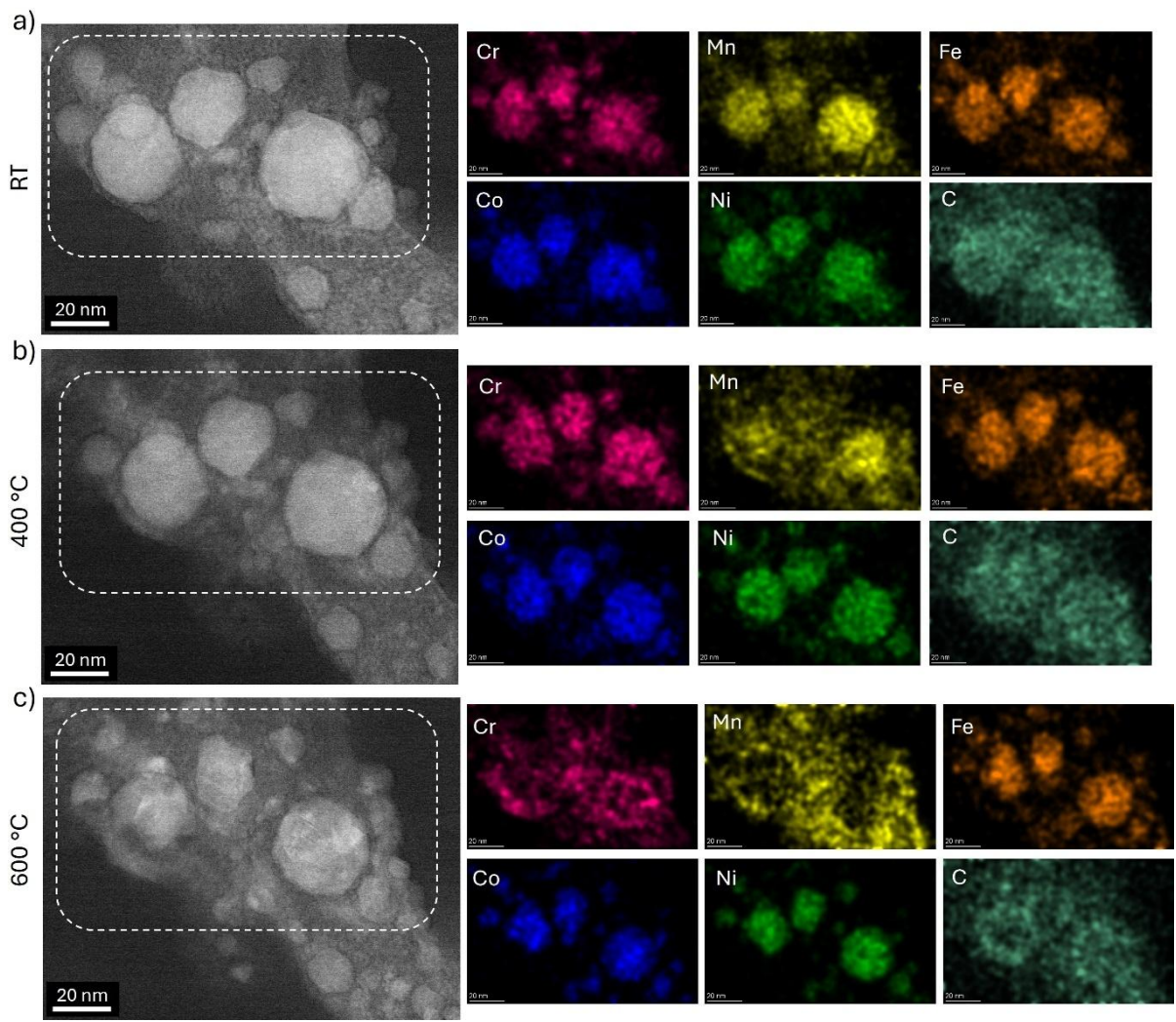

Figure S3 a), b), c) STEM dark field image and corresponding EDS mappings of the constituent elements of the HEA NPs synthesized in acetone at RT, annealed at 400 °C and 600 °C, respectively.

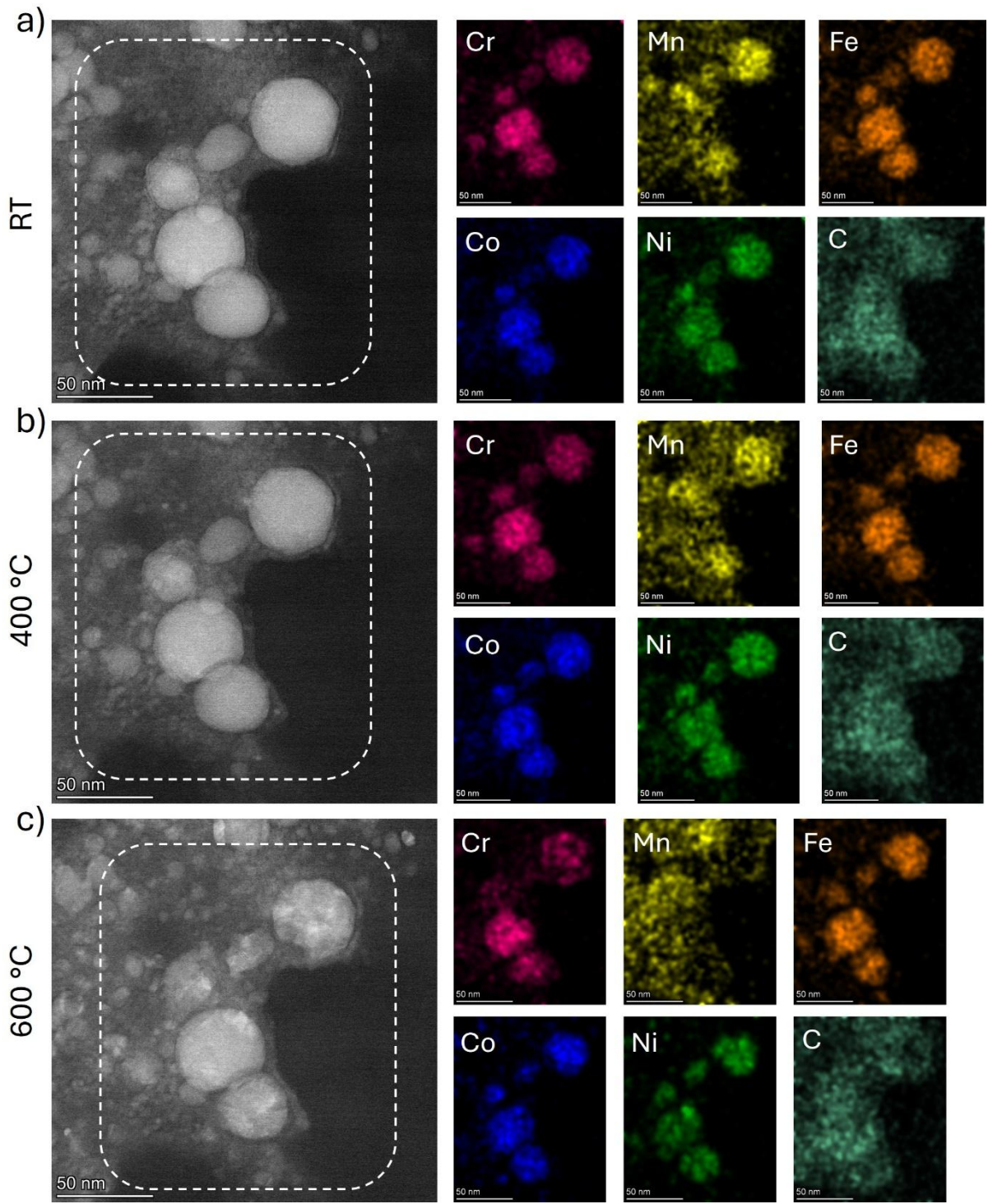

Figure S4 a), b), c) STEM dark field image and corresponding EDS mappings of the constituent elements of the HEA NPs synthesized in acetone at RT, annealed at 400 °C and 600 °C, respectively.

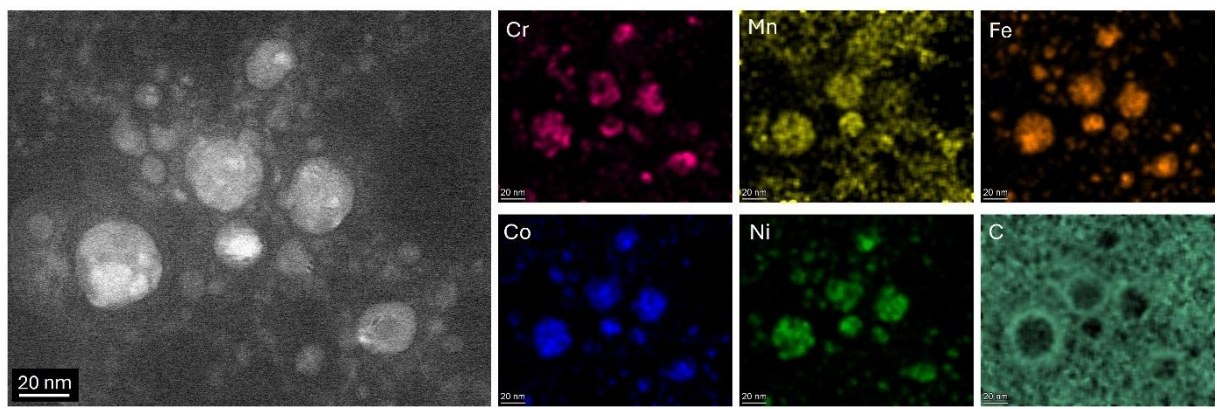

Figure S5 STEM-EDS mappings of HEA NPs synthesized in acetonitrile annealed at 600 °C.

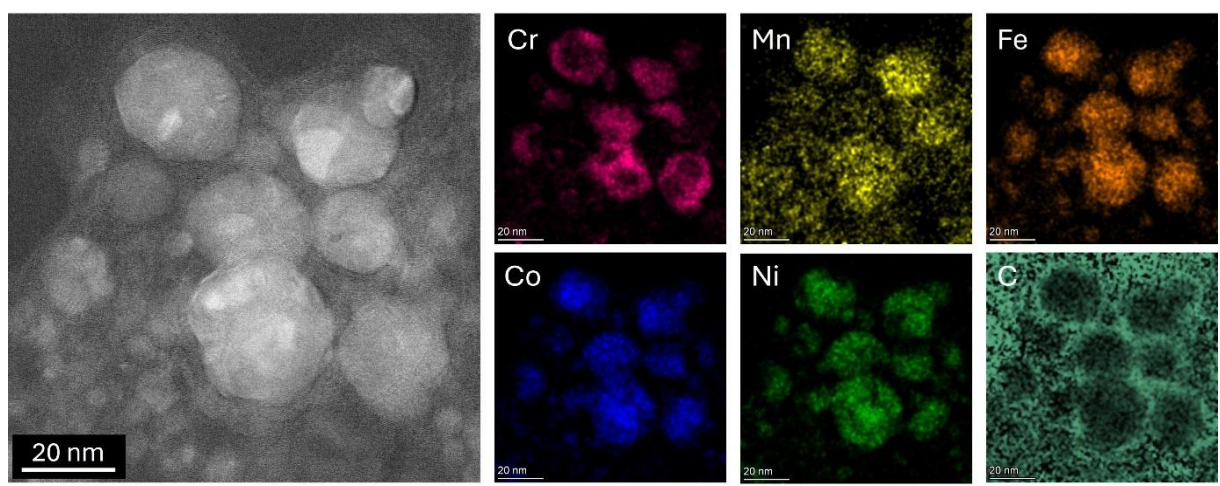

Figure S6 STEM-EDS mappings of HEA NPs synthesized in acetonitrile annealed at 600 °C.

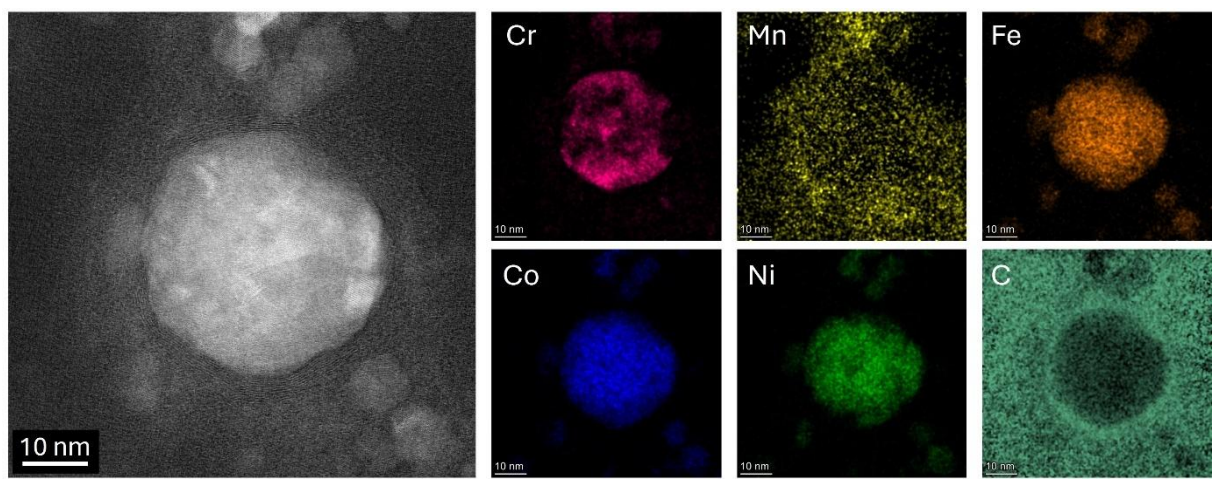

Figure S7 STEM-EDS mappings of HEA NPs synthesized in acetonitrile annealed at 600 °C.

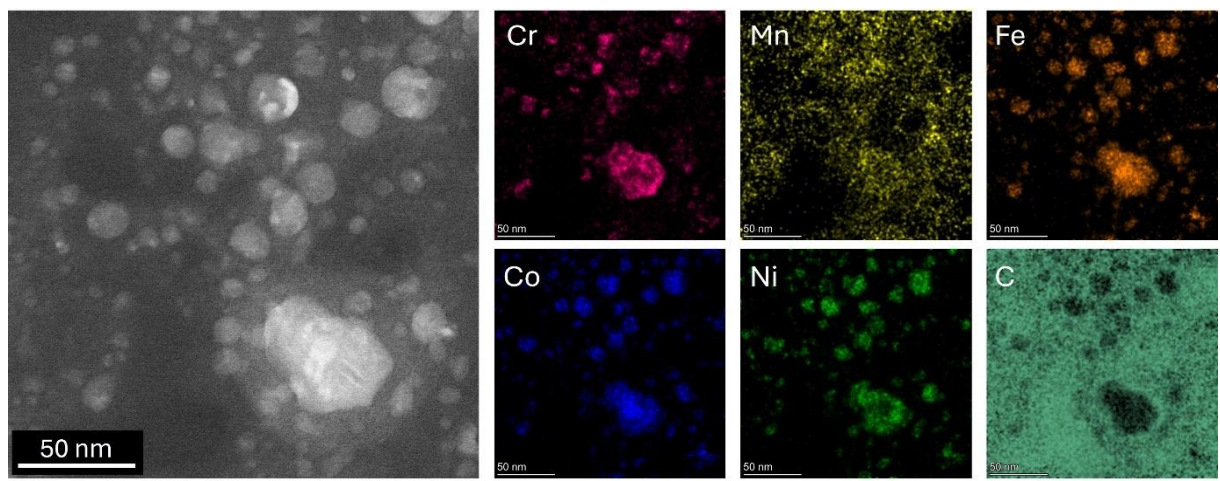

Figure S8 STEM-EDS mappings of HEA NPs synthesized in acetonitrile annealed at 600 °C.

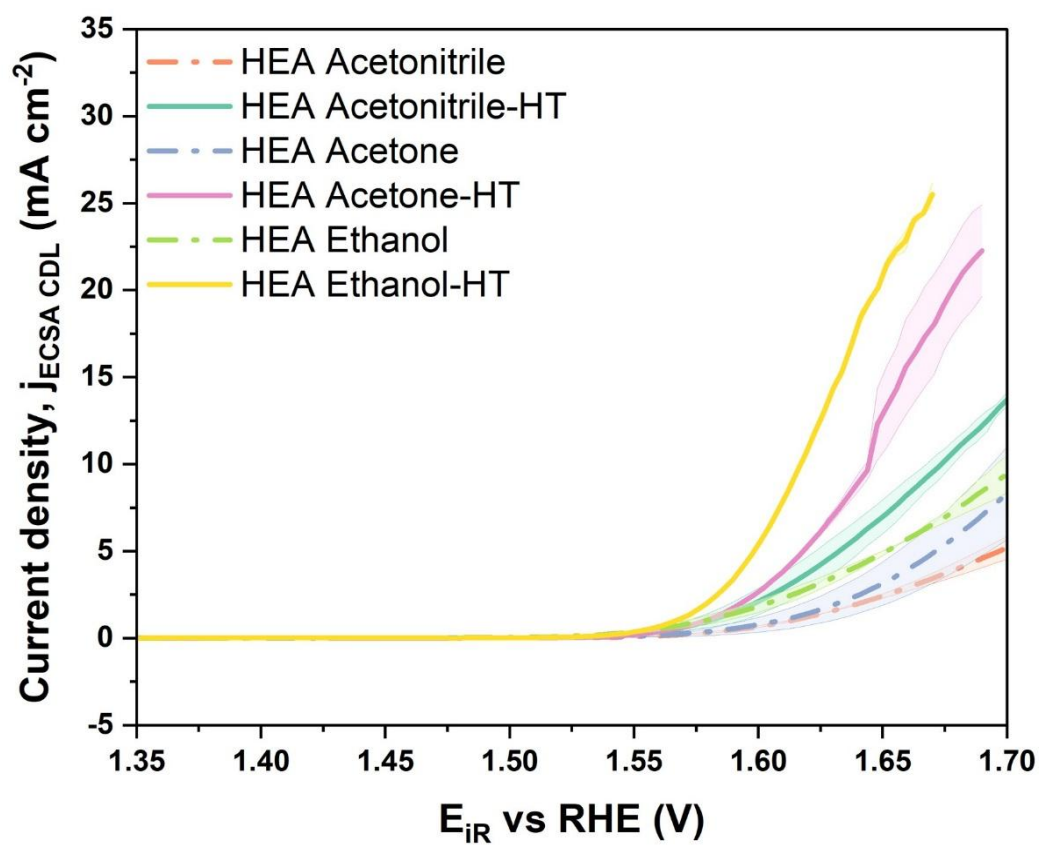

Figure S9 Linear sweep voltametry curves (current density normalized with ECSA) of the as-synthesized and annealed (denoted as HT) HEA NPs synthesized in acetonitrile, acetone and ethanol, respectively.

The EIS data were fitted using the equivalent electric circuit (EEC) shown in the inset of Figure S10. The model consists of the components arranged in $R_s + Q_{dl} / (R_{ct} + C_a + R_a)$, where R1 or $R_s$ is the solution resistance, Q1 or $Q_{dl}$ corresponds to the double layer capacitance. R2 or $R_{ct}$ represents the charge transfer resistance. C1 ($C_a$) and R3 ($R_a$) represent the adsorption capacitance and adsorption resistance. In this model constant phase element (Q) is used instead of capacitance in order to count for the non-ideal behavior. The coefficient *a* reflects the deviation from the ideal behavior [55,60].

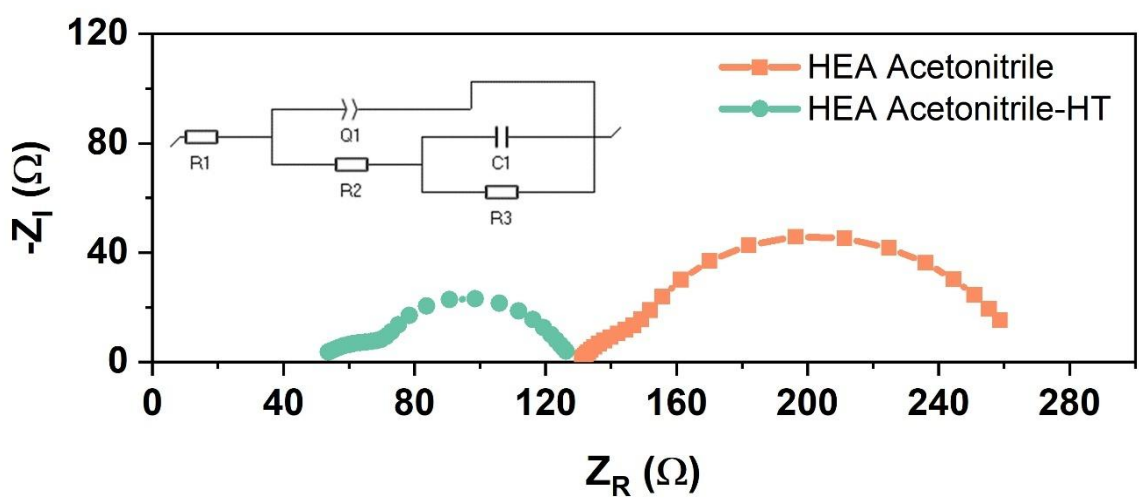

Figure S10 Nyquist plots for the as-synthesized and annealed HEA NPs synthesized in acetonitrile with the equivalent circuit for the spectrum given in the inset. The experiment was carried out at 1.6 V in 0.1 M KOH. Frequency range: 1Hz – 30kHz with 10 mV amplitude and 6 step dec$^{-1}$.

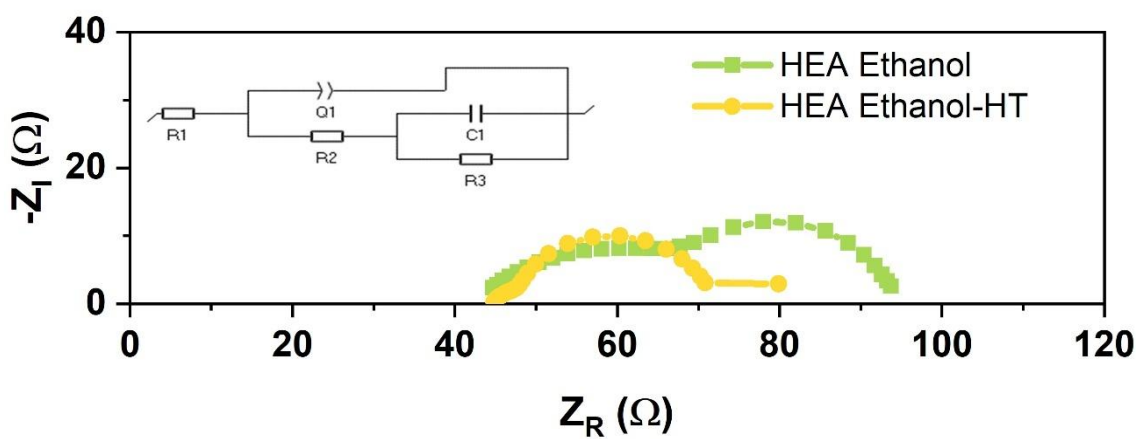

Figure S11 Nyquist plots for the as-synthesized and annealed HEA NPs synthesized in ethanol with the equivalent circuit for the spectrum given in the inset. The experiment was carried out

at 1.6 V in 0.1 M KOH. Frequency range: 1Hz – 30kHz with 10 mV amplitude and 6 step dec[-1].

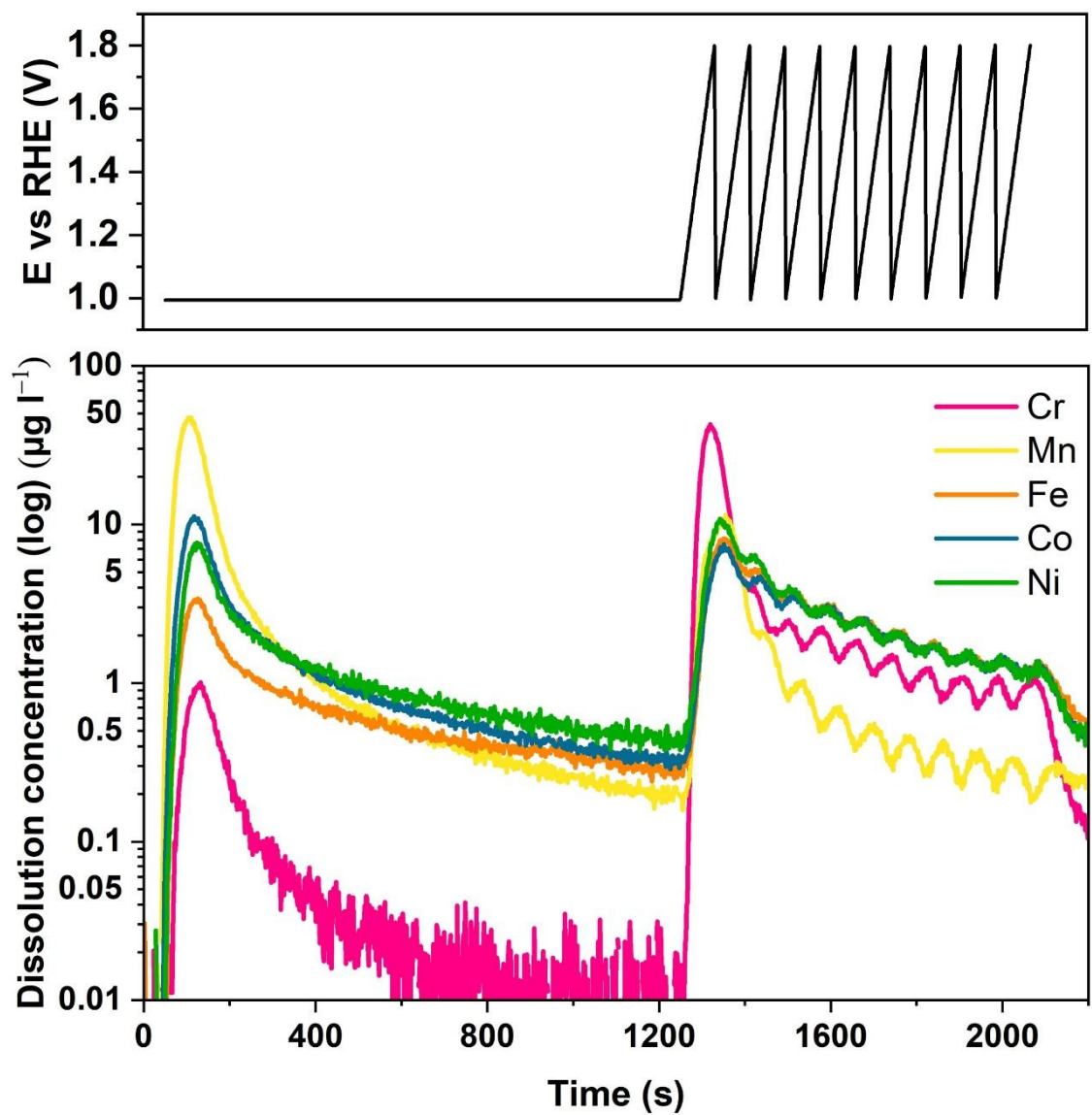

Figure S12 Applied potential and the corresponding online ICP-MS dissolution profiles of dissolved metal ions as a function of time of the annealed HEA NPs in acetonitrile.

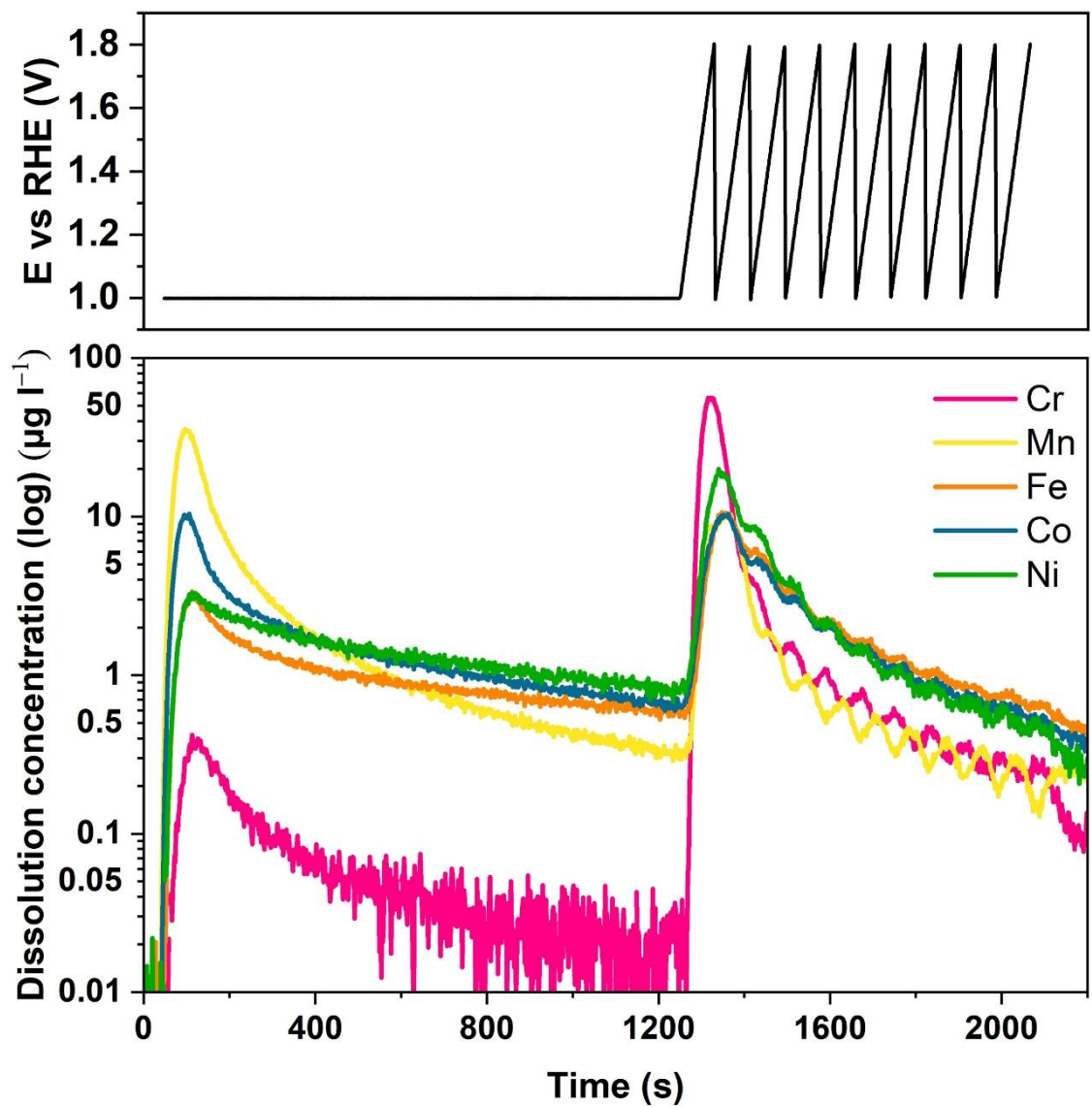

Figure S13 Applied potential and the corresponding online ICP-MS dissolution profiles of dissolved metal ions as a function of time of the annealed HEA NPs in ethanol.

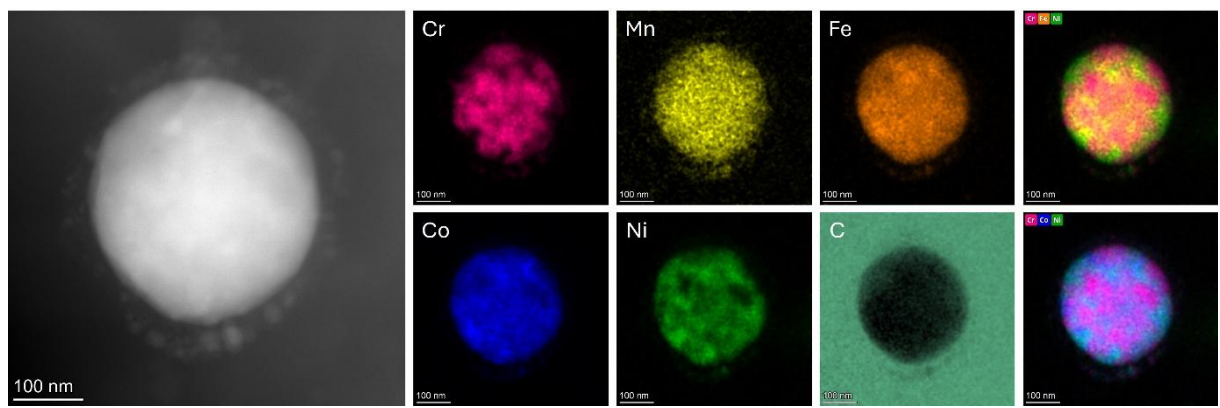

Figure S14 STEM-EDS mapping and corresponding Cr-Fe-Ni and Cr-Co-Ni elemental mapping overlays of an annealed HEA NP (~320 nm in diameter) synthesized in acetone after OER cycling, highlighting the retention of the heterostructured microstructure after electrochemical testing.

Table S1 Constraints used during the peak fitting of XPS 3p spectra listed for each element.

|  | Cr 3p | | Mn 3p | | Fe 3p | | Co 3p_#1 | |
|---|---|---|---|---|---|---|---|---|
|  | min | max | min | max | min | max | min | max |
| Pos. Constr. / eV | 46 | 47 | 47 | 50.4 | 53.7 | 56.7 | 59.9 | 62.5 |
| FWHM Constr. / eV | 0.6 | 5 | 0.6 | 5 | 0.6 | 5 | 0.6 | 5 |

|  | Co 3p_#2 | | Ni 3p | |
|---|---|---|---|---|
|  | min | max | min | max |
| Pos. Constr. / eV | 59.8 | 64.5 | 67.5 | 69 |
| FWHM Constr. / eV | 0.6 | 2 | 0.6 | 5 |

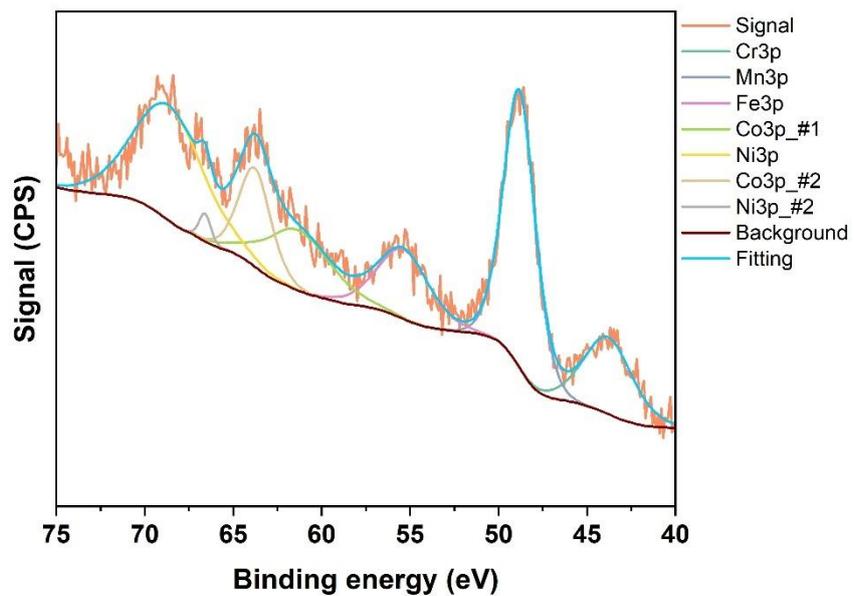

Figure S15 Deconvoluted 3p XPS spectra of as-synthesized HEA NPs in acetonitrile.

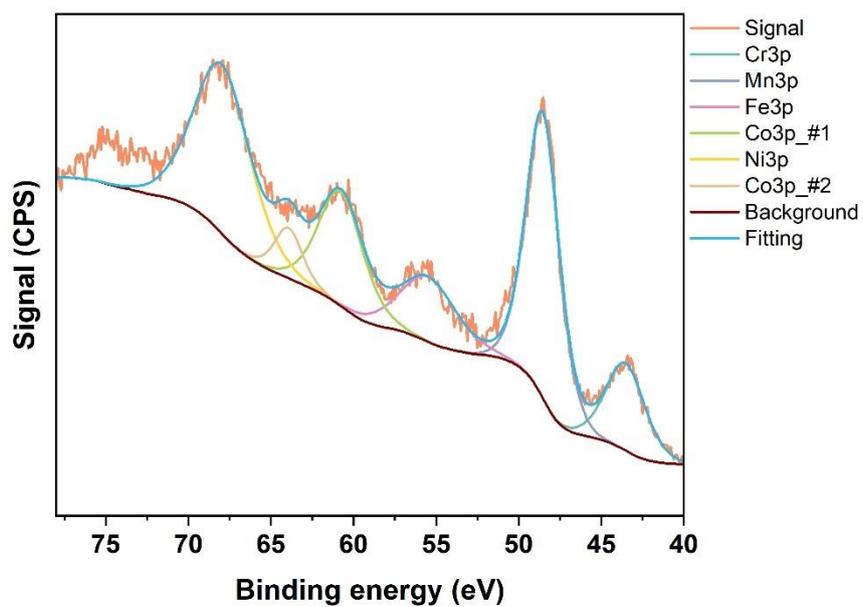

Figure S16 Deconvoluted 3p XPS spectra of as-synthesized HEA NPs in acetone.

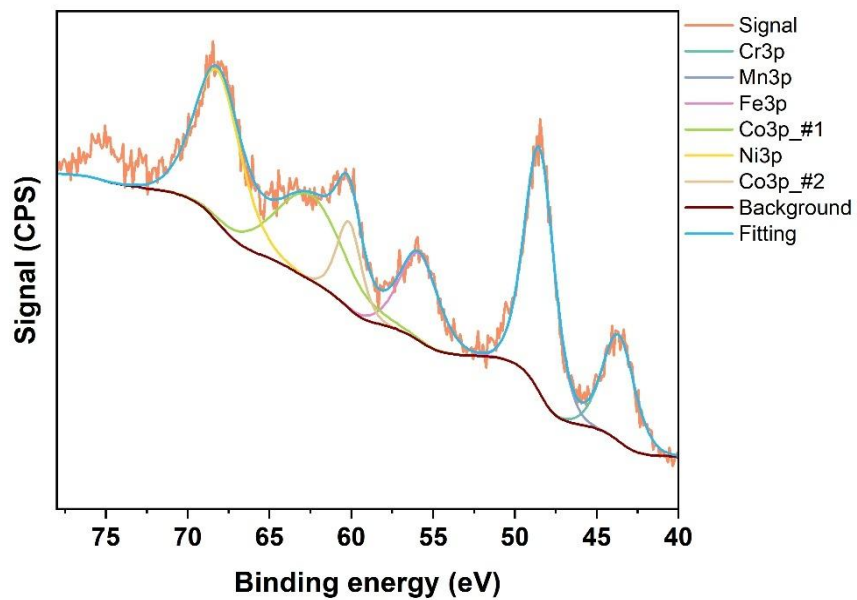

Figure S17 Deconvoluted 3p XPS spectra of as-synthesized HEA NPs in ethanol.

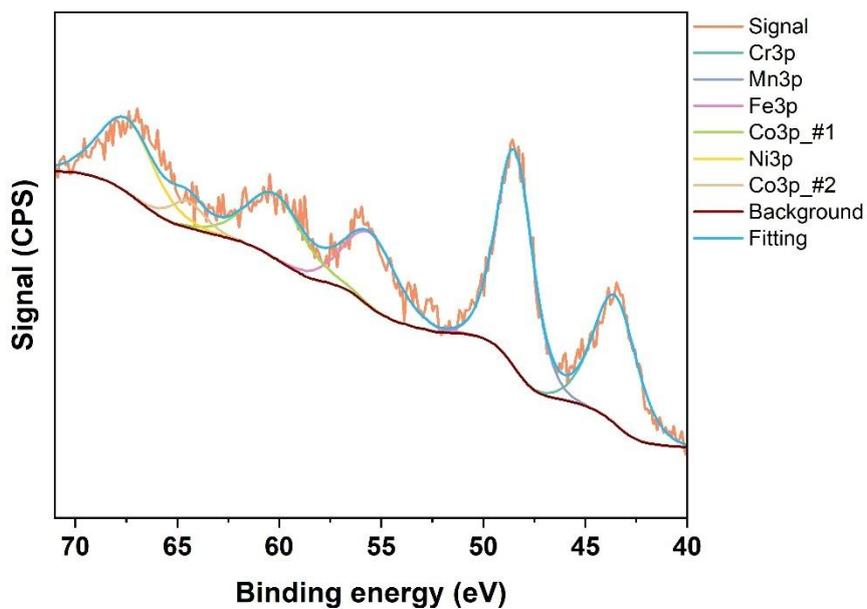

Figure S18 Deconvoluted 3p XPS spectra of annealed HEA NPs in acetonitrile.

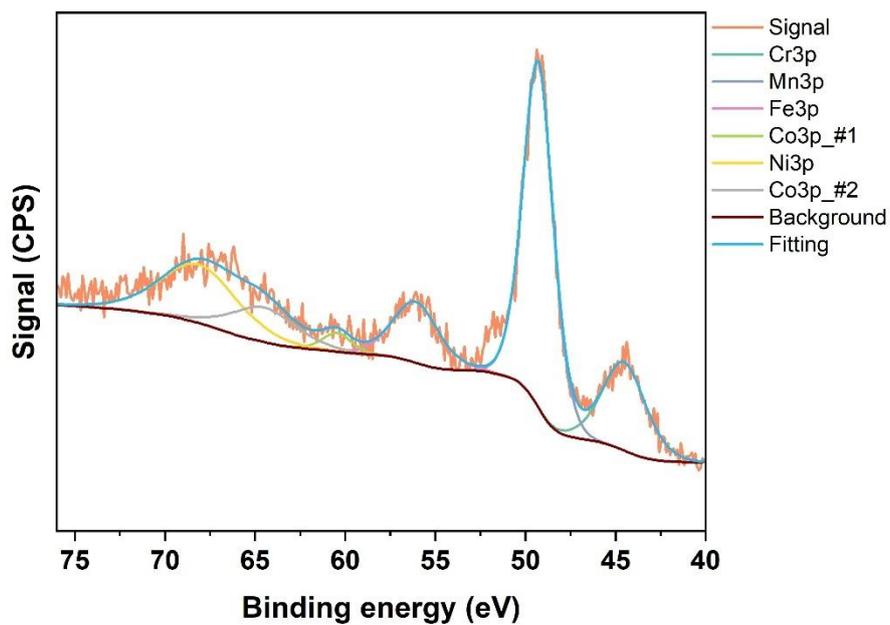

Figure S 19 Deconvoluted 3p XPS spectra of annealed HEA NPs in acetone.

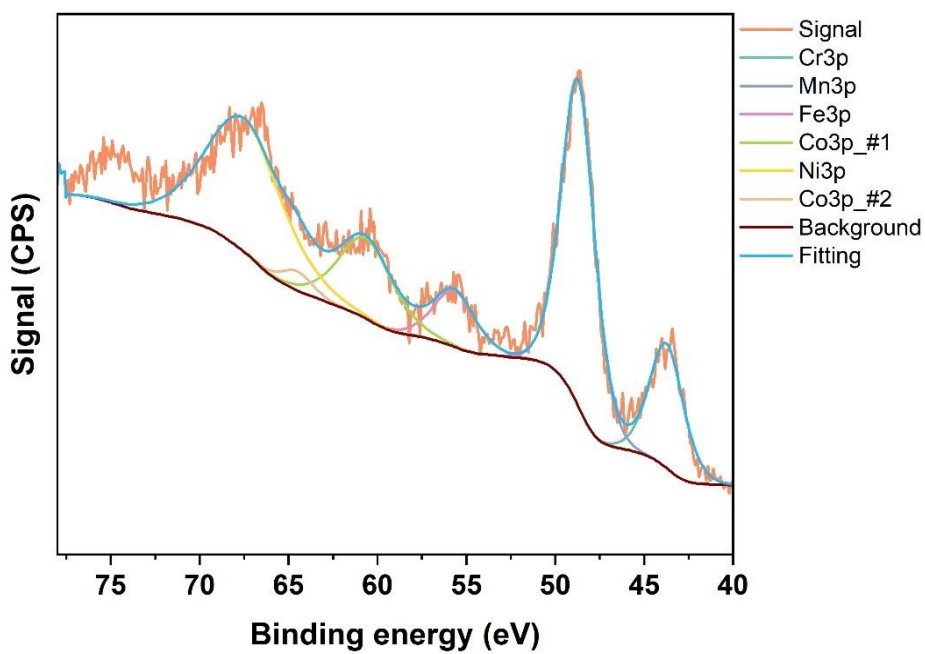

Figure S20 Deconvoluted 3p XPS spectra of annealed HEA NPs in ethanol.

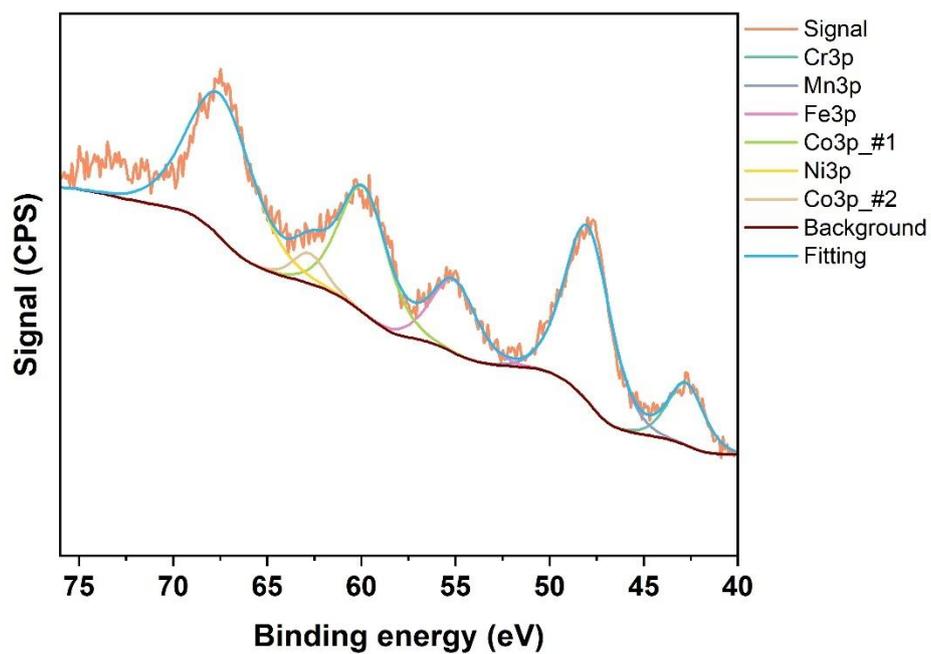

Figure S21 Deconvoluted 3p XPS spectra of annealed and electrochemically tested HEA NPs in acetonitrile.

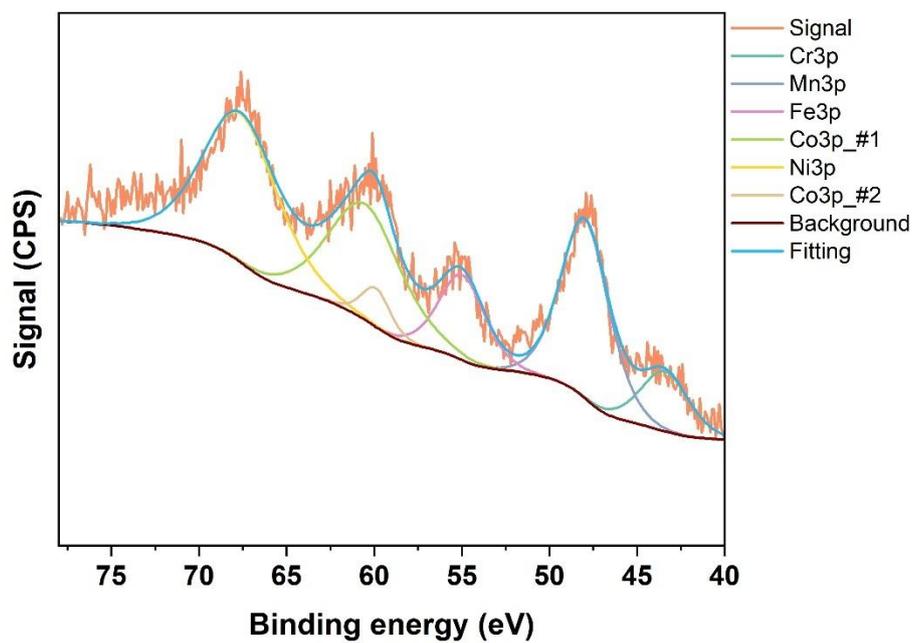

Figure S22 Deconvoluted 3p XPS spectra of annealed and electrochemically tested HEA NPs in acetone.

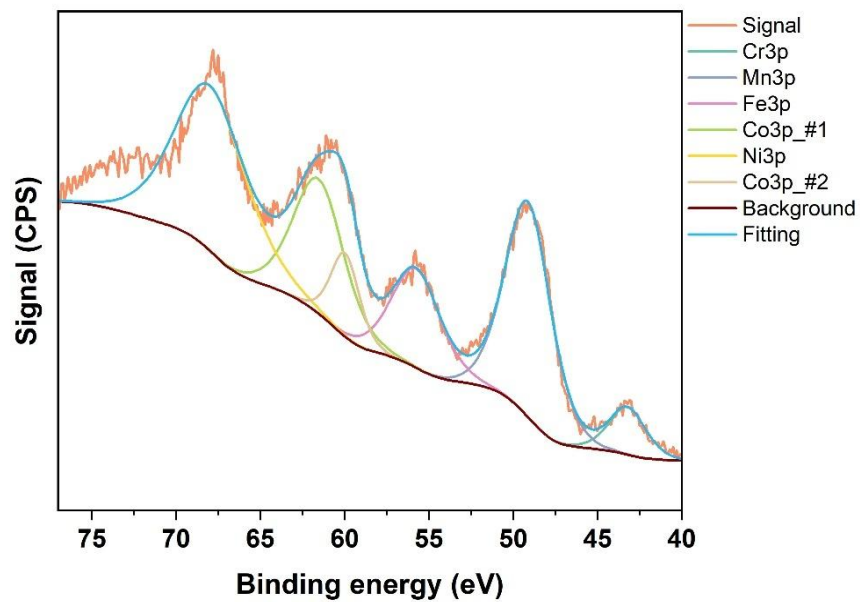

Figure S23 Deconvoluted 3p XPS spectra of annealed and electrochemically tested HEA NPs in ethanol.